\documentclass[12pt,showpacs,amssymb,amsmath,floatfix,preprint]{aastex}
\usepackage{graphicx}
\def\beq{\begin{eqnarray}}
\def\eeq{\end{eqnarray}}
\def\bsp{\begin{split}}
\def\esp{\end{split}}

\def\d{\mathrm{d}}

\def\wM{\widetilde{M}}
\newcommand{\mbold}[1]{\mbox{\boldmath{\ensuremath{#1}}}}
\begin{document}

\title{Perihelion Precession in spherically symmetric Kaluza-Klein Theories}
\author{{\O}ystein Elgar{\o}y}
\affil{Institute of theoretical astrophysics, University of Oslo, PO Box 1029 Blindern, 
\\N-0315 Oslo, Norway}
\email{oystein.elgaroy@astro.uio.no}
\author{{\O}yvind Gr{\o}n\altaffilmark{1}}
\affil{Oslo College, Department of Engineering, Cort Adeleres gt. 30, N-0254 Oslo, Norway}
\email{Oyvind.Gron@iu.hio.no}
\and
\author{Sigbj{\o}rn Hervik\altaffilmark{2}}
\affil{Department of Mathematics and Statistics, Dalhousie University, Halifax, Nova Scotia, B3H 3J5 Canda}
\email{herviks@mathstat.dal.ca}
\altaffiltext{1}{Department of Physics, University of Oslo, PO Box 1048 Blindern, N-0316 Oslo, Norway}
\altaffiltext{2}{Address after 15 April, 2008: Department of Mathematics and Natural Sciences, University of Stavanger, N-4036 Stavanger, Norway}

\date{\today }

\begin{abstract}
We consider the perihelion precession and bending of light in a class of Kaluza-Klein models and show that the ``electric redshift'' model, proposed in Zhang (2006) to explain the redshift of Quasars, does not agree with observations. As Zhang's model only considers the Jordan frame, we also compute the perihelion precession as seen in the Einstein frame and show that, to lowest order, the perihelion precession can only be consistent with observation in the Einstein frame. However, when we consider the corresponding ``electric redshift'' in the Einstein frame, we find that the redshift is significantly lower than for the Jordan frame and is therefore very unlikely to explain the redshift of Quasars.
\end{abstract}
\keywords{gravitation --- relativity --- quasars:general}


\section{Introduction} 
In this Letter we will consider a 5-dimensional Kaluza-Klein (KK) model for our Universe and check whether this model is consistent with the classical tests of the general theory of relativity. The KK models (see, e.g., \citet{KK}) consider a small extra dimension of space which has a remarkable effect as seen from a 4-dimensional point-of-view. The effect of this extra dimension enables us to view electromagnetism and gravity in 4 spacetime dimensions as a single 5 dimensional theory of gravity. The size of the 4th extra spatial dimension is directly related to the electromagnetic field strength. 

Treating the size of the extra dimension as dynamical, we introduce, as seen from a 4-dimensional perspective, naturally a scalar field associated with the size of the extra dimension. More specifically, consider a 5-dimensional metric: 
\beq
\d s_{KK}^2=g_{\mu\nu}\d x^\mu \d x^\nu+\Phi^2\left(\d y+A_{\mu}\d x^\mu\right)^2.
\eeq
In KK theory, the extra dimension is spanned by ${\mbold\xi}=\partial/\partial y$, and, from a 4 dimensional interpretation, $A_{\mu}$ is the electromagnetic vector-potential, and $\Phi$ is a scalar field. 

There have been many investigations regarding the physical consequences of KK theory. It was pointed out by \citet{ADD} that we have not measured the gravitational force law at distances much less than a tenth of a millimetre, so if there exist curled up dimensions that only affect gravity, they could be as large as the minimum allowed by such measurements. This would solve the so-called ``hierarchy of scales problem'', namely that the energy scale of the standard model physics is many orders of magnitude smaller that the Planck energy scale where gravity is expected to unify with the other fundamental forces \citep{GRW,HLZ,HS,HKHA}.

In \citet{Zhang} it was proposed that such a KK theory could explain the redshift of quasars. In particular, the author considered a class of exact 5 dimensional solutions to KK theory originally derived by \citet{CD}. The solutions considered were spherically symmetric and are therefore of great interest also for models for our solar system. The redshift due to the extra dimension, which was termed 'electric redshift', was suggested to explain the large redshift of quasars. 

Here, we will consider the same models and derive the perihelion precession and the bending of light in these models. We perform the calculation for both the  Jordan frame and the Einstein frame (explained later). As we will see, the found perihelion precession and bending of light is consistent with the observations only in the Einstein frame (in \citet{Zhang}  only the Jordan frame is considered).

\section{Weak-field approximation}
Let us consider a 4-dimensional spherically symmetric metric which in the weak-field approximation can be approximated as (in isotropic coordinates)
\beq
\d s^2&=&-\left(1-\frac{2\wM}{r}\right)\d t^2\nonumber \\ 
&&+\left(1+\frac{2M}{r}\right)\left(\d r^2+r^2\d \theta^2+r^2\sin^2\theta\d\phi^2\right).
\label{eq:weakfield}\eeq
This metric encompasses many known metrics, for example, using the Schwarzschild metric, and expanding to 1st order in $(M/r)$ gives $\wM=M$. This special case therefore gives the standard perihelion precession given in textbooks. 

\subsection{Perihelion precession}
Let us find the perihelion precession for the metric (\ref{eq:weakfield}) assuming the following: 
\begin{itemize}
\item{} $(M/r)$ and $(\wM/r)$ are small (so that we can expand in terms of these parameters). 
\item{} Non-relativistic velocities: Test particle is moving with velocity so that $v/c\ll 1$.  
\end{itemize} 
We will later apply the result to Solar system planets in which case these assumptions are valid to great accuracy. 

We introduce the proper time of the particle, $\tau$, and consider the Lagrangian 
\[ L=\frac 12\left(\frac{\d s}{\d \tau}\right)^2\]
so that the geodesic equations are equivalent to the Euler-Lagrange equations
\[ \frac{\d}{\d\tau}\left(\frac{\partial L}{\partial \dot{q}^i}\right)-\frac{\partial L}{\partial q^i}=0, \] along with the four-velocity identity $v_{\mu}v^{\mu}=-1$. 
Furthermore, due to the spherical symmetry there is no loss of generality in assuming that the particle moves in the plane $\theta=\pi/2$, say.

The coordinates $t$ and $\phi$ are both cyclic which leads to the two constants of motion: 
\beq
p_{t}&=& \left(1-\frac{2\wM}{r}\right)\dot{t}, \nonumber \\
p_{\phi}&=& \left(1+\frac{2M}{r}\right)r^2\dot{\phi}. \nonumber
\eeq
The four-velocity identity now gives us an eqution for $\dot{r}$. It is convenient to consider $r$ as a function of $\phi$; i.e., $r(\phi)$. Using the standard trick $\dot{r}=(\d\phi/\d\tau)(\d r/\d\phi)$, and differentiating w.r.t. $\phi$, we get the orbit equation
\beq
u''+u=\frac{\widehat M}{p_{\phi}^2}+3Mu^2,
\label{ueq}\eeq
where ${}'\equiv\d/\d\phi$, $u\equiv 1/r$ and 
\[\widehat{M}\equiv(\wM-M)p_t^2+M.\]
Note that $\widehat{M}$ corresponds to the Newtonian mass. 

To lowest order the solution to this equation is
\beq
u_0=\frac 1p(1+e\cos\phi),
\eeq
where $p=a(1-e^2)=p_\phi^2/\widehat{M}$, $a$ is the semi-major axis, and $e$ is the eccentricity of the orbit. Furthermore, assuming $v/c\ll 1$ we can use the four-velocity identity to estimate $p_t^2$ to be $p_t^2=1+\mathcal{O}(\widehat{M}/a)$, and hence, 
\[ \widehat{M}=\wM[1+\mathcal{O}(\wM/a)]. \]  
We can interpret this as follows: To lowest order, the orbit is an ellipse and the test particle orbits a mass given by $\widehat{M}\approx \wM$. 

The second term in eq.(\ref{ueq}) gives the precession of the orbit and using the standard method \citep{Book}, we can calulate the precession angular velocity to be
\beq
\omega_p=\frac{3\widehat{M}^{\frac 32}}{a^{\frac 52}(1-e^2)}\left(\frac{M}{\widehat{M}}\right).
\eeq
\subsection{Bending of light}

Let us also consider the bending of light. Using a similar calculation we can find the orbit equation for photons to be: 
\beq
u''+u=\frac{(\wM-M)p_t^2}{p_{\phi}^2}+3Mu^2.
\label{ueqlight}
\eeq
In this case the geodesics are light-like so $v_{\mu}v^{\mu}=0$ which we can use to estimate $p_t^2$. If $b$ is the impact parameter, we get at the closest point $p_t^2\approx p_{\phi}^2/b^2$. 

Let us now try the ansatz: 
\beq
u=\frac 1b\left(\cos\phi+B+A\sin^2\phi\right). 
\eeq
To first order in $\wM/b$ and $M/b$ we get
\[ A=\frac{M}{b}, \quad B=\frac{\wM}{b},\] 
which determines the constants $A$ and $B$. The bending of the photon path  as it passes the massive body is now found by solving $u\left(\pi/2+\delta\phi/2\right)=0$. This gives
\beq
\delta\phi=2(A+B)=\frac{2(\wM+M)}{b}.
\eeq
\section{5-dimensional spherically symmetric Kaluza-Klein model} 
25 years ago Chodos and Detweiler considered a spherically symmetric solutions to 5-dimensional general relativity \citep{CD}. In fact, they were able to find exact solutions to the vacuum equations and interpreted the solutions for the various ranges of the parameter values. 

Here, we will not be needing the exact solutions but we will consider the weak field approximation and find a simpler, and more illuminating, expression. We therefore consider the weak field approximation for a 5-dimensional spherically symmetric spacetime, and assume the 5-dimensional metric to be \citep{CD}: 
\beq
\d s^2_{KK}&=&-\left(1-\frac{2M_1}{r}\right)\d t^2+\frac{2A}{r}\d t\d x+\left(1+\frac{2B}{r}\right)\d x^2\nonumber \\ &+&\left(1+\frac{2M_2}{r}\right)\left(\d r^2+r^2\d \theta^2+r^2\sin^2\theta\d\phi^2\right).
\label{eq:KKweakfield}\eeq
This metric is asymptotically flat and spherically symmetric. The extra dimension corresponds to the coordinate $x$. This metric is  correct to order $1/r$. The Ricci tensor contains second derivatives of the metric, hence, we must assume that the Ricci tensor must be correct to order $(1/r)^3$. 

Assuming the metric satisfies the 5D vacuum equations the term $(1/r)^3$ gives 
\[B=M_1-M_2.\]
The parameter $A$ only contributes to order $(1/r)^4$, so the above choice gives $R_{\mu\nu}=\mathcal{O}(1/r^4)$ as desired. 

The metric (\ref{eq:KKweakfield}) can, from a 4-dimensional point of view, be interpreted as a 4-dimensional spacetime with an electromagnetic field, and a scalar field:
\beq
\Phi^2=g_{55}=\left(1+\frac{2B}{r}\right), \quad B=M_1-M_2.
\label{eq:scalar}\eeq
Here, $g_{55}$ is the metric component associated with the 5th dimension, and $\Phi$ is the scalar field. 

The 4-dimensional spacetime is commonly interpreted in two different frames; namely, the Jordan and Einstein frame.  A long outstanding question has been which of these frames is the 'most physical'. The question is related to how we interpret the various frames. The various frames are \emph{mathematically equivalent}; however, their physics may be different depending on how one wants to interpret the frame. The Jordan and the Einstein frames are related via a conformal transformation, and if we choose to let our measuring rods and clocks be mutable and adjustable and independent of the theory, then these frames will also be \emph{physically equivalent}. If, on the other hand, we choose to use fixed conventions, using, for example, the corresponding metric to measure distances, then only one frame can be the correct one. If such fixed conventions are used, the two frames are \emph{physically inequivalent} (for a discussion on this issue, see \citet{Flanagan}). Here, we will insist on using the corresponding metrics as the ones that determines the geodesics of test particles. In such cicumstances the physics may be different in the different frames. Setting aside the question which frame is the most physical one, we will determine perihelion precession and bending of light in both frames under the above assumptions. 

\paragraph{Jordan Frame:} 
In the Jordan frame, the 4-dimensional spacetime, $\d s_J^2$ in the weak field approximation, is given by (\ref{eq:weakfield}) with $\wM=M_1$ and $M=M_2$. Therefore, by measuring the orbital period of planets, the Newtonian mass is approximately given by $M_1$. 
\paragraph{Einstein Frame:} 
The Einstein frame is conformally related to the Jordan frame as follows: 
\[ \d s_{E}^2=\Phi \d s_{J}^2.\] 
The scalar field, $\Phi$, is given by eq.(\ref{eq:scalar}), so the 4-dimensional $tt$- and $rr$-components of the metric are 
\beq 
g_{tt}=-\Phi\left(1-\frac{2M_1}{r}\right)&\approx & -\left(1-\frac{M_1+M_2}{r}\right), \nonumber \\
g_{rr}=\Phi\left(1+\frac{2M_2}{r}\right)&\approx & \left(1+\frac{M_1+M_2}{r}\right). \nonumber
\eeq
Thus in the weak field approximation, eq.(\ref{eq:weakfield}), $\wM=M=(M_1+M_2)/2$. Here, the Newtonian mass is interpreted as $(M_1+M_2)/2$. 

\section{Zhangs model} 
In \citet{Zhang}, Zhang considered spherically symmetric Kaluza-Klein models and suggested that the redshift of an electrically charged object with density and mass comparable to that of a neutron star, can  give rise to a redshift similar to that of a quasar. 

Zhang's model is a particular C-D solution, which in the weak-field approximation gives: 
\[ M_1=M_2(2+3\alpha^2), \quad M_2=\frac{2Gm}{3\sqrt{1+\alpha^2}}.\]
Here, $\alpha$ is taken to be a free parameter of the theory.  
This implies that, in the Jordan frame, the perihelion precession, and bending of light, are: 
\[ \omega_p=\omega_{pE}\left(\frac{1}{2+3\alpha^2}\right), \quad \delta\phi=\delta \phi_E\left[\frac{3(1+\alpha^2)}{2(2+3\alpha^2)}\right]. \] 
Here, the index $E$ indicates the prediction given by ordinary Einstein gravity. 
It is therefore clear that the perihelion precession will give significant deviation for the perihelion precession of planetary orbits regardless of the value of $\alpha$. 

For the Einstein frame, Zhang's model gives
\[ \omega_p=\omega_{pE}, \quad \delta\phi=\delta\phi_{E}.\] 
In this case, the weak-field approximation gives, to lowest order, the same results as for regular Einstein gravity. To lowest order, these tests are therefore compatible with observations. We must therefore consider higher-order corrections to be able to distinguish the KK-model and ordinary Einstein gravity. 

In the Einstein frame, the 4-dimensional metric is, to second order (here, $M=\sqrt{3}(1+\alpha^2)B$): 
\beq
g_{tt}&=&-1+\frac{2M}{r}-\frac{2(1+3\alpha^2)M^2}{(1+\alpha^2)r^2}+\cdots \\
g_{rr}&=&1+\frac{2M}{r}-\frac{2(-2+3\alpha^4)M^2}{3(1+\alpha^2)^2r^2}+\cdots 
\label{2ndorder}\eeq
In the PPN formalism \citep{Will}, we have 
\beq
g_{tt}=-1+2U-2\beta U^2, \quad g_{rr}=1+2\gamma U+\cdots, 
\eeq
where $U=M/r$, and $\gamma$ and $\beta$ are so-called PPN parameters. In the Einstein frame we see that $\gamma=1$ which, as pointed out, agrees with experiments. Furthermore, regarding the PPN parameter $\beta$ (which measures the mount of non-linearity in the superposition law of gravity), experiments involving perihelion observations and also the Nordtvedt effect put stong constraints on its value\citep{Will}: $|\beta -1| \leq 2.3\times 10^{-4}$. Therefore, the value of $\beta$ must be close to unity. 

From the expressions of Zhang's model in the Einstein frame to second order, eq.(\ref{2ndorder}), we see that a value of $\beta$ close to 1 is only compatible with a small value of $\alpha$. 
Hence, we are lead to the conclusion that \emph{only in the Einstein frame, and for small values of the parameter $\alpha$, are the models by Zhang compatible with the measurements of perihelion precession and bending of light.}. 

\begin{figure}
  \includegraphics[width=8cm,angle=-90]{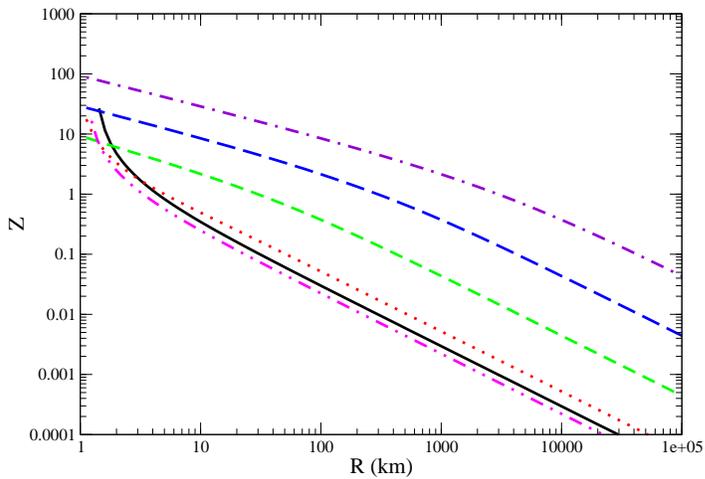}
\caption{The ``electric redshift'' of Zhang and the redshift from a Reissner-Nordstr{\"o}m (RN) solution as a function of the radial coordinate in isotropic coordinates ($R$). The RN solution has charge $Q=10^{20}\textrm{C}$ and mass $M=1.5M_{\mathrm{Sun}}$, and corresponds to the the pink, dot-dot-dashed line. The other lines correspond to Zhang's ``electric redshift'' with $M=1.5M_{\mathrm{Sun}}$, and the parameter $\alpha$ as follows: black, solid line, $\alpha=0$; red, dotted line, $\alpha=1$; green, short-dashed line, $\alpha=10$; blue, long-dashed line, $\alpha=100$; purple, dot-dashed line, $\alpha=1000$. }
\label{Zhang-RN}\end{figure}

\section{Redshift in the Einstein frame} 
Now, all calculations done of the redshift in \citet{Zhang} are in the Jordan frame. However, as we have seen, this description of the KK models is not compatible with the perihelion precession and bending of light. It is therefore necessary to consider these redshifts in the Einstein frame.

\begin{figure}
  \includegraphics[width=8cm,angle=-90]{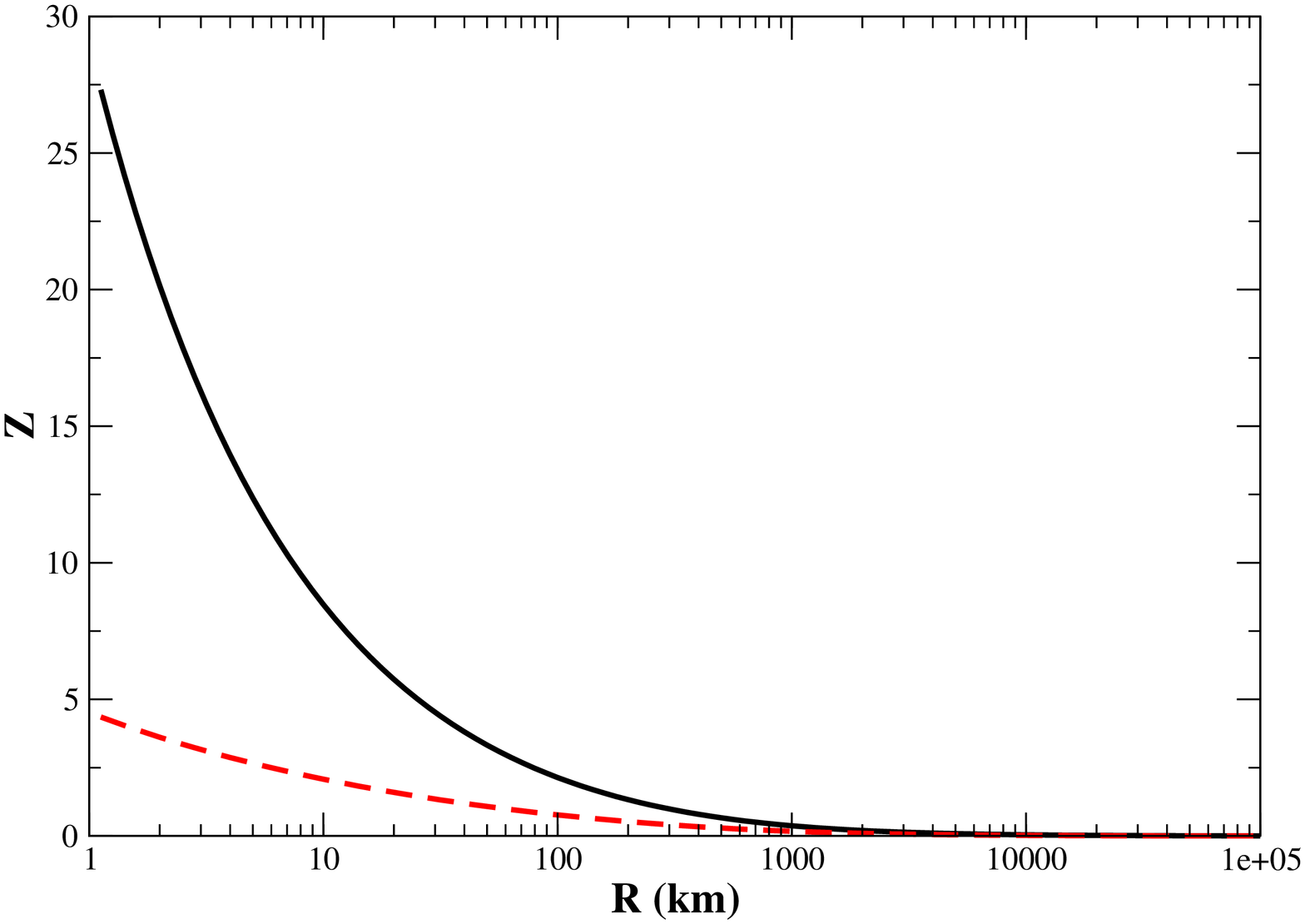} \\
\includegraphics[width=8cm,angle=-90]{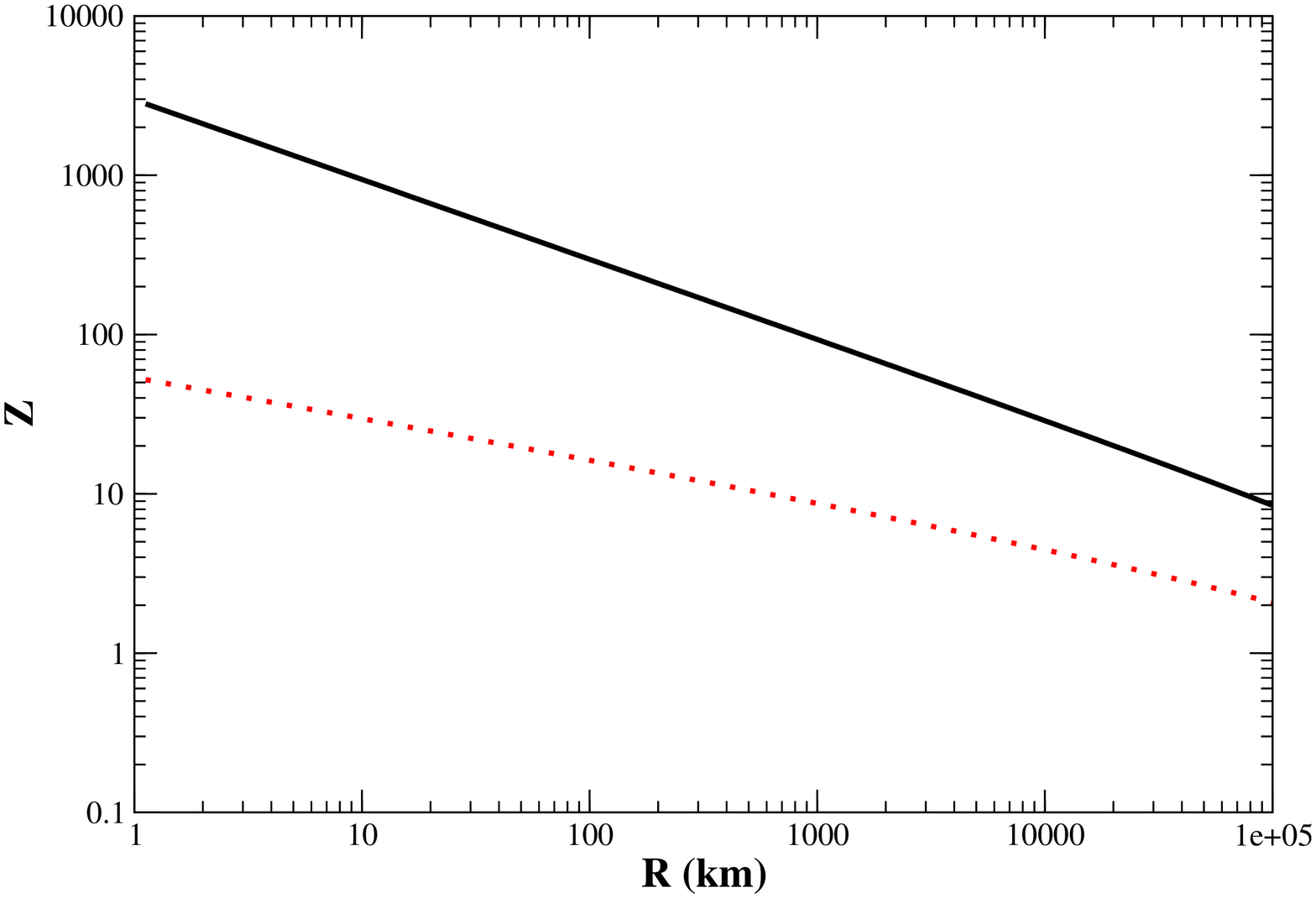}
  \caption{Einstein vs Jordan redshift: Here, the redshift as seen in the Einstein and the Jordan plotted for comparison. The solid black line is the Jordan frame redshift, while the dashed red line is the redshift in the Einstein frame. For these figures the values of $\alpha$ are $100$ (top) and $10^6$ (bottom). One can clearly see that the redshift in the Einstein frame is significantly smaller than in the Jordan frame. }\label{EvsJ}
\end{figure}

First, in Figure \ref{Zhang-RN} we have plotted Zhang's ``electric redshift'' as a function of the radial coordinate of the set of isotropic coordinates. This plot therefore correspond the redshift in Zhang's paper \citep{Zhang} and shows the redshift in the Jordan frame. For comparison, in the same figure, we have also plotted the redshift from a Reissner-Nordstr{\"o}m (RN) star. We can see that RN star gives a lower redshift than the ``electric redshift'' of Zhang's.   

In Figure \ref{EvsJ} the redshift, $z$, is plotted as a function of the radial coordinate in isotropic coordinates for two values of the parameter $\alpha$. As we can clearly see, the redshift in the Einstein frame is significantly lower than in the Jordan frame. This fact is even more evident for higher values of $\alpha$. It is still possible to get sufficiently large redshifts consistent with observation, however, this requires significantly larger values of $\alpha$ than in the Jordan frame. However, as pointed out, the PPN formalism puts very tight bounds on the value of $\beta$ and, in effect, requires $\alpha$ to be small. The values of $\alpha$ required  to explain the redshift of Quasars gives $\beta\approx 3$ which seems to contradict the experimental value of $\beta\approx 1$. 

\section{Discussion}
The real breakthrough of the general theory of relativity was its capability of explaining and predicting several features of the gravitational interaction between stars, planets, and light. One of these features was a perihelion precession of the Mercurian orbit which Newtonian gravity left unexplained. The perihelion precession of Mercury, along with bending of light, are therefore considered among the pillars of the general theory of relativity. 

As we have seen, the model proposed in \citet{Zhang} which was aimed to explain the high redshift of light emitted by Quasars, does not fit well with the observed perihelion precession. In fact, the model of Zhang predicts a precession less than a half of that of the observed value of the Mercurian orbit. As the Zhang considered only the Jordan frame, we consequently used the Einstein frame for which the perihelion precession is, to lowest order in the weak field limit,  consistent with observations. However, in order for the corresponding redshift to agree with the high redshift of Quasars, the parameter $\alpha$ has to be significantly larger than in the Jordan frame. This seems highly unlikely as a higher value of $\alpha$ induces pathologies for the 5-dimensional solution \citep{CD}. Furthermore, since the parameter $\alpha$ is related to the charge of the Quasar, the large values of $\alpha$ would indicate a extremely charged object which is, from a 4-dimensional point of view, known to be unstable. However, the most devastaing blow to this model is the fact that experiments indicate a value of the PPN parameter $\beta$ close to 1. This is inconsistent with a large value of $\alpha$; hence, is therefore very unlikely that the ``electric redshift'' is a viable model for the redshift of Quasars.

\acknowledgements 
{\O}E acknowledges support from the Research Council of Norway, Project No. 162830.


\begin{thebibliography}{}
\bibitem[Appelquist, Chodos, \& Freund(1987)]{KK} 
Appelquist, T., Chodos, A., \& Freund, P.G.O. 1987, \textit{Modern Kaluza-Klein Theories}, {Addison-Wesley} 

\bibitem[Arkani-Hamed, Dimopoulos, \& Dvali(1998)]{ADD} Arkani-Hamed, N.,  Dimopoulos, S., \& Dvali, G. 1998, Phys.Lett. B, 429, 263 

\bibitem[Giudice, Rattazzi, \& Wells(1999)]{GRW} Giudice, G. F.,  Rattazzi, R., \& Wells, J. D. 1999,  
Nucl. Phys. B, 544, 3

\bibitem[Han, Lykken, \& Zhang(1999)]{HLZ} Han, T., Lykken, J. D., \& Zhang, R. J. 1999, Phys. Rev. 
D, 59, 105006
\bibitem[Hewett \& Spiropulu(2002)]{HS} Hewett, J., \& Spiropulu, M. 2002, Ann. Rev. Nucl. Part. Sci., 
52, 397 .

\bibitem[Hoyle et al.(2004)]{HKHA} Hoyle, C. D.,  Kapner, D. J., Heckel B. R., Adelberger E. G.,  
Gundlach, J. H.,  Schmidt, U., Swanson, H. E. 2004,  Phys. Rev.D,  70, 042004

\bibitem[Zhang(2006)]{Zhang}
Zhang, T.X. 2006, \apjl, 636,  L61


\bibitem[Chodos \& Detweiler(1982)]{CD} 
Chodos, A., \& Detweiler, S. 1982, Gen. Rel. Grav., 14, 879

\bibitem[Gr{\o}n \& Hervik(2007)]{Book}
Gr{\o}n, {\O}.,  \& Hervik, S. 2007,  \textit{Einstein's General Theory of Relativity}, Springer: New York 

\bibitem[Flanagan(2004)]{Flanagan}
Flanagan, E. E. 2004,  Class. Quant. Grav., 21, 3817

\bibitem[Will(1993)]{Will}
Will, C. M. 1993,  \textit{Theory and Experiment in Gravitational Physics} Cambridge University Press; 
C.M. Will, \textit{The Confrontation between General Relativity and Experiment} (2006), \texttt{http://www.livingreviews.org/lrr-2006-3}  
\end{thebibliography}
\end{document}